\documentclass[aps,prb,preprint,showpacs,groupedaddress]{revtex4}

\usepackage{amsmath}
\usepackage{amssymb}
\usepackage{graphicx}
\usepackage{dcolumn}
\usepackage{bm}
\usepackage{subfigure}

\begin{document}

\title{Ground-state properties and superfluidity of two- and   
quasi two-dimensional solid $^4$He}

\author{C. Cazorla$^{1}$, G. E. Astrakharchik$^{2}$, J. 
Casulleras$^{2}$, and J. Boronat$^{2}$}

\affiliation{
$^{\rm 1}$ Department of Chemistry, University College London, London WC1H 0AH, UK \\
$^{\rm 2}$ Departament de F\'\i sica i Enginyeria Nuclear, Campus Nord
B4-B5, Universitat Polit\`ecnica de Catalunya, E-08034 Barcelona, Spain}

\email{c.silva@ucl.ac.uk}

\begin{abstract}

In a recent study we have reported a new type of trial wave function
symmetric under the exchange of particles and which is able to describe 
a supersolid phase.
In this work, we use the diffusion Monte Carlo method and this 
model wave function to study the properties of solid $^4$He in two- and 
quasi two-dimensional geometries. 
In the purely two-dimensional case, we obtain results for the total 
ground-state energy and freezing and melting densities which are in 
good agreement with previous exact Monte Carlo calculations performed
with a slightly different interatomic potential model. 
We calculate the value of the zero-temperature superfluid fraction 
$\rho_{s} / \rho$ of 2D solid $^{4}$He and find that it is negligible 
in all the considered cases, similarly to what is obtained in the perfect 
(free of defects) three-dimensional crystal using the same computational approach.
Interestingly, by allowing the atoms to move locally in the perpendicular direction 
to the plane where they are confined to zero-point oscillations 
(quasi two-dimensional crystal) we observe the emergence of a finite superfluid 
density that coexists with the periodicity of the system. 

\end{abstract}


\maketitle

\section{Introduction}
\label{sec:introduction}

Quantum crystals are characterized by unusually large atomic kinetic energy,  
Lindemann ratio and non-negligible anharmonicity even at low temperatures
and high pressures.~\cite{cazorla09,cazorla08a}
The counterintuitive possibility of simultaneous solid order and
superfluidity in solid $^4$He, the most representative of quantum 
crystals, has long attracted the interest of both theoreticians
and experimentalists. After several
unfruitful experiments to detect superfluid signals in solid $^4$He, Kim
and Chan reported few years ago the first evidence of non-classical rotational
inertia (NCRI) both in confined environment~\cite{moses2} and in
bulk.~\cite{moses1} From then on,
several other experimental groups (up to $5$, so far) have observed NCRI using different
samples containing small or ultra small $^3$He concentrations, in a simple
crystal or in a polycrystal, and using several annealing
schemes.~\cite{balibar08} There is almost overall 
agreement of all the data concerning the onset temperature 
$T_0=75-150$~mK at which the superfluid fraction becomes zero, the lowest
value corresponding to ultra pure samples (only 1 ppb $^3$He).
However, the experimental values of the superfluid density
reported so far change by more than one order of magnitude ($\rho_s/\rho \simeq 0.03-0.5$\%)
depending on the purity, annealing conditions in which the crystal is grown, etc.   
Such high dispersion suggests that the superfluid signal observed in solid $^4$He 
is probably due to the presence of some defects in
the crystal, which could be of different nature: dislocations, vacancies or
grain boundaries.~\cite{sasaki} 

On the theoretical side, path integral Monte Carlo (PIMC) calculations performed  
at low temperatures (down to $0.1$~K) show that perfect commensurate $^{4}$He crystal does posses 
neither finite superfluid fraction~\cite{ceper1} nor condensate fraction.~\cite{ceper2} 
Only non-zero $\rho_s/\rho$ has been estimated in the presence of
disorder introduced in the form of a glassy phase~\cite{glass} and of defects like 
dislocation lines,~\cite{dislocation} and vacancies.~\cite{cazorla08} 
Moreover, our recent calculations based on the diffusion 
Monte Carlo (DMC) method show negligible superfluid fraction of perfect bulk solid $^4$He
at strictly zero temperature.~\cite{cazorla08}
These DMC calculations have been performed using a new model of trial wave
function which allows simultaneously for spatial solid order and Bose-Einstein
symmetry and with the benefit of a simple use for importance sampling. 
It has been shown that the energetic and structural
properties of solid $^{4}$He can be reproduced very accurately with this trial
wave function model when it is used for importance sampling in the diffusion
Monte Carlo method.

In this work, we extend our study of solid helium to purely two-
and quasi two-dimensional geometries relying upon similar computational 
approaches to the ones used in Refs.~[\onlinecite{cazorla08,cazorla08b}].  
The motivation for carrying out the present study is four-fold. 
First, relevance of quantum fluctuations is generally enhanced in systems of 
reduced dimensionality hence possible signatures of superfluidity may be 
detected more easily. Second, from a computational point of view
low-dimensional systems are reasonably affordable  
so one can explore wide thermodynamic ranges on them. 
Third, in a recent study~\cite{cazorla08b} of strictly two-dimensional solid 
H$_{2}$ we have shown that when the density of particles
is reduced down to practically the spinodal point a finite superfluid fraction  
is observed to appear in the film; in this work we carry out similar investigations 
on solid $^{4}$He (that is, in the metastable regime) in order to unravel possible connections 
between the density of particles and superfluid fraction.   
And fourth, theoretical predictions on low-dimensional model 
systems can provide valuable understanding on experimental
realization of solid helium confined to restricted geometries and 
also on interpretation of supersolid signatures in 
general.~\cite{gordillo09,pierce99,greywall91,corboz08} 

There is previous work done on the estimation of the ground-state properties
of strictly two-dimensional solid $^{4}$He.  
Many years ago, Whitlock \emph{et al.}~\cite{gfmc} performed a systematic 
study of the energetic and structural properties of this system based on the Green's 
function Monte Carlo approach (GFMC). Posteriorly, Gordillo 
\emph{et al.}~\cite{gordillo98} estimated the phase diagram of two-dimensional
$^{4}$He over a range of temperatures and coverages using PIMC calculations. 
More recently, various authors have reported on  
the melting transition and dynamical properties of helium films using variational and 
\emph{exact} ground-state methods.~\cite{shadowmelting,shadow2} 
Interestingly, Vitali \emph{et al.}~\cite{shadow2} 
have investigated the existence of off-diagonal long range order (ODLRO) in 
2D solid $^{4}$He using the zero-temperature version of PIMC adapted to the shadow wave 
function formalism, the so-called shadow path integral ground-state method (SPIGS). 
In the present article, we provide comparison with respect to the results reported in 
these previous works and present new predictions as well.  
In particular, we report on direct estimations of the superfluid fraction in 2D solid 
helium and its dependence on the density of particles. We also analyse 
the superfluid behavior of a quasi two-dimensional crystal, i. e. an ensemble of He atoms  
confined within a plane but allowed to explore the out-of-plane direction locally,
and assess its dependence on the density of particles and degree of confinement. 
In the quasi two-dimensional case, we observe the presence of a superfluid signal which 
coexists with the periodicity of the system.  
 
The remaninder of the article is organized as follows. In Section~\ref{sec:method},
we summarize the basics of the DMC method and describe the symmetrized trial wave function 
model used throughout this work. Next, we report  
results for the ground-state properties of two- and quasi two-dimensional solid helium. 
Finally, we present some discussion and the conclusions  
in Section~\ref{sec:discussion}.

\section{Method and trial wave function}
\label{sec:method}

We study the ground-state of two-dimensional solid $^4$He by means of the DMC
method and Hamiltonian
$H= -\hbar^2/2m_{\rm He} \sum_{i=1}^{N} \nabla_i^2 + \sum_{i<j}^{N} V(r_{ij})$~,
with $N$ being the number of particles. The He-He atomic interaction 
is modelled with the semi-empirical pairwise potential due to Aziz \emph{et al.}~\cite{aziz2}
(heretofore referred to as Aziz~II).
The DMC method solves stochastically the imaginary-time ($\tau$) Schr\"odinger equation
providing essentially exact results for the ground-state energy and diagonal properties
of bosonic systems within controllable statistical errors.
For $\tau \to \infty$, sets of configurations (walkers) $\bf{R}_i\equiv\{\bf{r}_1,\ldots,\bf{r}_N\}$ 
generated with DMC render the probability distribution function ($\Psi_0 \Psi$), where $\Psi_0$ and
$\Psi$ are the ground-state wave function and trial wave function for
importance sampling, respectively. The short-time Green's function approximation
that we use, and according to which the walkers evolve, is accurate up to order $(\Delta \tau)^3$;
technical parameters in the calculations, as for instance the mean population of
walkers ($= 400$) and time step $\Delta \tau$ ($= 5 \cdot 10^{-4}$~K$^{-1}$), have been adjusted in order 
to eliminate possible bias in the total energy per particle to less than $0.02$~K/atom.~\cite{boro,casulleras} 

Customarily, structural and energetic properties of solid $^4$He are 
explored with the Nosanow-Jastrow trial wave function 
\begin{equation}
\Psi_{\rm NJ}({\bf r}_1,\ldots,{\bf r}_N) = \prod_{i<j}^{N} f(r_{ij}) 
\prod_{i,I=1}^{N} g(r_{iI})= \psi_{J} \psi_{L}~,
\label{njtrial}
\end{equation}  
where $N$ is the number of particles (in this work we consider commensurate crystals
only so $N$ is also equal to the number of lattice sites), 
$f(r)$ being a two-body correlation factor accounting for atomic 
correlations, and $g(r)$ a one-body localization factor which accounts for
the periodicity in the system by linking every 
particle to a particular lattice site of a perfect crystal structure. 
The wave function $\Psi_{\rm NJ}$ leads to an excellent
description of the equation of state and structural properties of quantum 
solids~\cite{cazorla}
but it can not be used to estimate properties which are directly related
to the quantum statistics. The reason of this is that  $\Psi_{\rm NJ}$ is not symmetric 
upon the exchange of particles and it misses the quantum statistics of the system. 

In a recent work,~\cite{cazorla08} we have introduced a new type of wave function,
 $\Psi_{\rm SNJ}$, which reproduces crystalline order and fulfills Bose-Einstein symmetry
requirements simultaneously. This model wave function is expressed as 
\begin{equation}
\Psi_{\rm SNJ}({\bf r}_1,\ldots,{\bf r}_N) = \prod_{i<j}^{N} f(r_{ij}) 
\prod_{J=1}^{N} \left( \sum_{i=1}^{N} g(r_{iJ}) \right)~,
\label{snjtrial}
\end{equation} 
where the product in the second term runs over lattice site indexes. 
Compact and managable analytical expressions for the drift velocity and
kinetic energy derive from Eq.~(\ref{snjtrial}), so  
$\Psi_{\rm SNJ}$ is very well-suited for implementation in DMC codes.
This model has proved to perform excellently in the description of bulk solid 
$^{4}$He and also p-${\rm H_{2}}$ in two dimensions.~\cite{cazorla08b} 
The key point in $\Psi_{\rm SNJ}$ is that the localization factor (second
term in Eq.~\ref{snjtrial})
is constructed in such a way that voids originated by multiple occupancy
of a same site are penalized (this feature will be 
illustrated in brief by a simple example). 

A similar model, $\Psi_{\rm LNJ}$, has been proposed recently by 
Zhai and Wu,~\cite{zhai} which is  
\begin{equation}
\Psi_{\rm LNJ}({\bf r}_1,\ldots,{\bf r}_N) = \prod_{i<j}^{N} f(r_{ij})
 \prod_{i=1}^{N} \left( \sum_{J=1}^{N} g(r_{iJ}) \right) 
\label{lpnjtrial}
\end{equation} 
and where the product in the second term runs over particle indexes.
This wave function also fulfills quantum symmetry requirements 
and is well-suited for DMC purposes, however it does not account for 
accurate description of quantum solids. 
We observe that when $\Psi_{\rm LNJ}$ is used for importance sampling 
solid order is not preserved but instead glassy-like configurations 
are generated in the simulations.~\cite{cazorla08b}
In fact, substantially better variational energies are obtained 
in two-dimensional solid hydrogen when using trial wave function
$\Psi_{\rm SNJ}$ instead of $\Psi_{\rm LNJ}$ 
(see Table~I in Reference~[\onlinecite{cazorla08b}]).
Similar variational outcomes are also found in two-dimensional 
solid $^{4}$He. For instance, at density $\rho = 0.525$~$\sigma^{-2}$ 
($\sigma = 2.556$~\AA) we obtain $E/N^{\rm SNJ} = 2.64~(4)$~K
($b = 1.1~\sigma$ and $a = 7.5~\sigma^{-2}$, see next section) whereas 
$E/N^{\rm LNJ} = 4.49~(15)$~K ($b = 1.3~\sigma$,  
$a = 7.5~\sigma^{-2}$ and $c = 4.0$, $c$ appearing in the exponent of 
the McMillan factor, see next section). 

\begin{figure}
\centerline{
        \includegraphics[width=0.8\linewidth,angle=0]{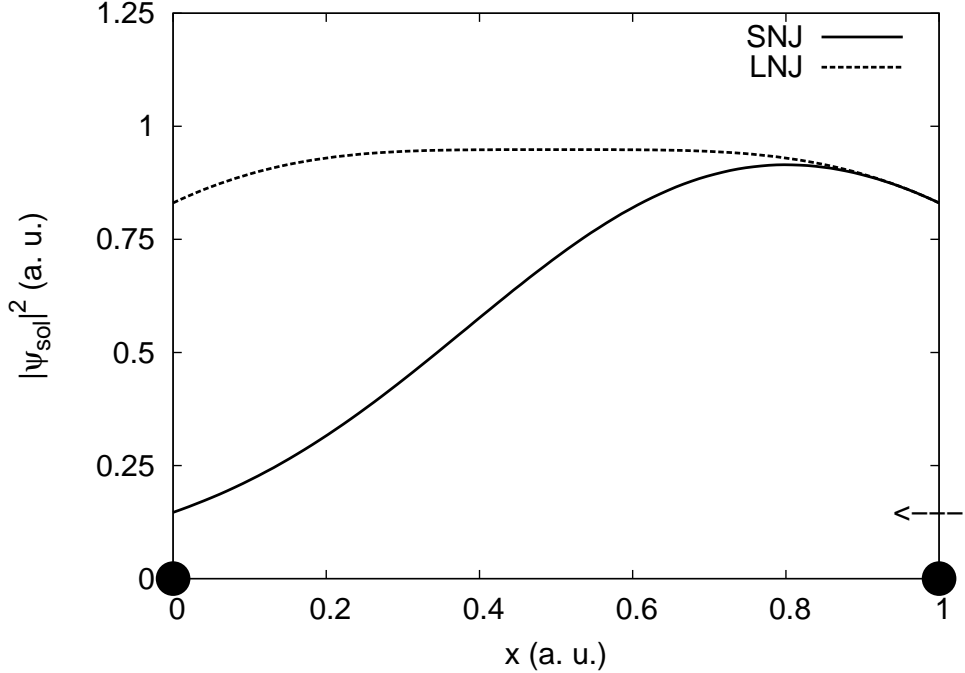}}%
        \caption{ $|\Psi_{\rm LNJ}|^{2}$ and $|\Psi_{\rm SNJ}|^{2}$ (Jastrow factor
	equal to unity) functions in the simple case of two particles moving in one dimension and
        with lattices sites separated by one arbitrary unity.}
\label{fig:symmetrized}
\end{figure}

The poor variational quality of wave function $\Psi_{\rm LNJ}$ can be 
understood in terms of the localization factor which, contrarily to 
what is required to keep solid order, does not penalizes multiple occupancy of a same site.
By multiple occupancy of a same site here we mean large probability of 
two particles near a same site to get too close one to the other 
(that is, as it would be allowed by the Jastrow factor alone).
Differences between trial wave function $\Psi_{\rm SNJ}$ and $\Psi_{\rm LNJ}$
can be illustrated by a simple example of two particles in a one-dimensional
lattice. For the sake of simplicity, we assume the distance 
between the atomic equilibrium positions to be one, the
parameter in the Gaussian factors ($g(r)$) $a = 1/2$ (in arbitrary units) 
and switch off the Jastrow factor. The value of the
square of wave function $\Psi_{\rm SNJ}$ and
$\Psi_{\rm LNJ}$, $|\Psi_{sol}|^{2}$, in the case of pinning one of the
particles in one lattice site (at $x=0$) and then move the second
particle towards it, is plotted in Fig.~\ref{fig:symmetrized}.  
As one observes in there, the value of
$\Psi_{\rm LNJ}$ at point $x = 1$ and $0$ (which correspond to particles  
placed over different sites and particles placed over the same position, respectively) 
is identical, whereas $\Psi_{\rm SNJ}(x = 1) > \Psi_{\rm SNJ}(x = 0)$~. 
Moreover, in the event of atomic overlap ($x = 0$) the drift force 
$1/\Psi \left( \partial \Psi / \partial x\right)$ corresponding to 
wave function $\Psi_{SNJ}$ is much more repulsive than that of wave function $\Psi_{LNJ}$.
In fact, the curve obtained in the $\Psi_{LNJ}$ case resembles that of a liquid where 
the localizing factor can be thought of a constant.
Also it must be noted that the value of $\Psi_{\rm LNJ}$ is maximum at half the way 
between $0$ and $1$, thus it will promote larger diffusion of the atoms throughout 
the volume. 

A symmetrized trial wave function that has been successfully applied to the study of solid
$^{4}$He is the called shadow wave function (SWF), proposed by 
Reatto \emph{et al.} more than twenty years ago.~\cite{vitiello88,reattoshadow} 
In the SWF formalism, an array of subsidiary particles (shadow particles) is  
defined and made to interact with the real atoms of the system; shadow particles are
correlated among them and their coordinates are integrated over the whole volume in 
such a way that bosonic symmetry requirements are fulfilled by construction. At the variational level 
the SWF has been shown to provide very accurate description of solid and liquid helium.
Nevertheless, this kind of trial wave function has never been used for importance sampling 
in a DMC calculation. In spite of this, very recently the SWF has been implemented within
the path integral ground state (PIGS) formalism so that variational
constraints in principle have been removed. This formalism has been used to 
explore solid $^4$He in two dimensions~\cite{shadow2} and
we will comment on their results in the next section. 

\section{Results}
\label{sec:results}

\subsection{2D solid $^{4}$He}
\label{subsec:2dhe4}

DMC simulations of 2D solid $^4$He in the triangular lattice configuration 
have been carried out for $N = 120$ particles in a $(x,y)$ box where periodic
boundary conditions are applied.
Correlation functions in Eq.~(\ref{snjtrial}) are chosen of McMillan,
 $f(r) = \exp\left[-1/2~(b/r)^{5}\right]$~, and  Gaussian, $g(r) = \exp\left[-1/2~(a r^{2})\right]$,
form.
Parameter $b$ and $a$ in factors $f(r)$ and $g(r)$  
have been optimized using variational Monte Carlo (VMC) at a density of 
$\rho = 0.480~\sigma^{-2}$ ($\sigma = 2.556$~\AA), and we obtain 
$b = 1.1~\sigma$ and $a = 7.5~\sigma^{-2}$ 
as best values (we neglect their weak dependence on density).
Size effects have been corrected by assuming that atoms distribute
uniformly beyond half the length of the simulation box.
Wave function $\Psi_{\rm SNJ}$ already provides a good description of
two-dimensional solid $^{4}$He at the variational level; 
for instance, at density $\rho = 0.550~\sigma^{-2}$ we obtain 
a variational total energy per atom $E/N$ of
$3.29~(3)$~K which must be confronted to the SWF result~\cite{shadowmelting} 
and variational benchmark $E/N^{SWF} = 3.10~(1)$~K, and the
Nosanow-Jastrow result $E/N^{NJ} = 3.18~(2)$~K.

\begin{figure}
\centerline{
        \includegraphics[width=0.8\linewidth,angle=0]{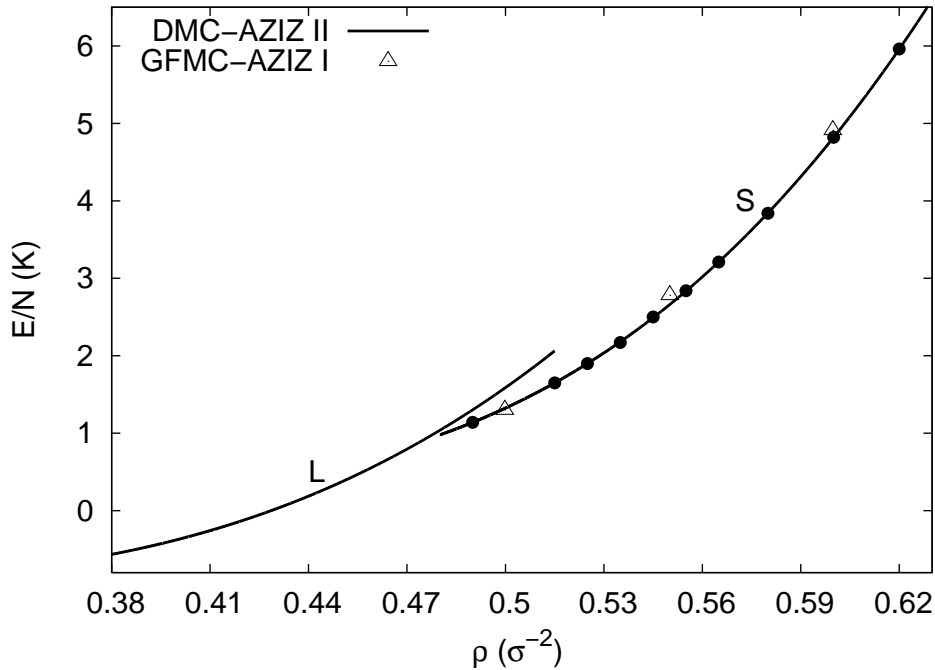}}%
        \caption{Energy per particle $E/N$ of solid (S) and liquid (L)
	   $^{4}$He in two dimensions and zero temperature. Previous
	   GFMC calculations~\cite{gfmc} obtained with the Aziz~I pair potential
	   are shown for comparison (triangles).
	   Results for the liquid phase are from Ref.~[\onlinecite{giorgini}].
	   }
\label{fig:eos}
\end{figure}

In Figure~\ref{fig:eos}, we plot DMC results for the total energy per particle $E/N$ as a 
function of density in solid and liquid $^{4}$He. Results for the
liquid phase are taken from Reference~[\onlinecite{giorgini}]. Previous
Green's function Monte Carlo (GFMC) estimations  
obtained with a slightly different atomic pairwise interaction  
than used here (referred to as Aziz I)~\cite{aziz} are included in the plot for comparison.~\cite{gfmc} 
In the solid phase, we fit our total energy per atom results to the polynomial curve 
$ E/N = e_{0} + a \rho + b \rho^{2} + c \rho^{3}$ with 
$e_{0} = -9.28(0.6)$~K, $a = 87.41(3.4)$~K$\sigma^{2}$, $b = -261.87(6.1)$~K$\sigma^{4}$ and 
$c = 258.91(3.7)$~K$\sigma^{6}$ the set of parameters which best reproduces them (statistical
uncertainties of the fit are expressed within the parentheses).  
Once the energy function $E(\rho)$ is known in the liquid and solid phases, 
the corresponding melting and freezing densities, namely $\rho_{l}$ and $\rho_{s}$, 
can be estimated by means of the double-tangent Maxwell construction. 
As a result, we obtain $\rho_{l} = 0.492~\sigma^{-2}$ and $\rho_{s} = 0.456~\sigma^{-2}$
which lie in between previous GFMC~\cite{gfmc} ($\rho_{l}^{\rm GFMC} = 0.471~\sigma^{-2}$~,
 $\rho_{s}^{\rm GFMC} = 0.443~\sigma^{-2}$) and variational SWF~\cite{shadowmelting} 
($\rho_{l}^{\rm SWF} = 0.522~\sigma^{-2}$, $\rho_{s}^{\rm SWF} = 0.475~\sigma^{-2}$) estimations.
The small discrepancy with respect to the GFMC results can be understood
in terms of the small differences in the atomic pairwise potential used. 
For instance, it is well-known that the Aziz~II potential provides atomic total energy  
values around $0.1$~K smaller than the Aziz~I potential does.~\cite{giorgini}

A well-known drawback of the NJ model (Eq.~\ref{njtrial}) 
is the impossibility of answering the fundamental question 
whether off-diagonal long range order (ODLRO) 
and/or superfluid behavior may manifest or not in quantum solids. The SNJ model 
(Eq.~\ref{snjtrial}) correctly fulfills the Bose-Einstein 
statistics and provides that information to some extent. 
Quantitatively, ODLRO is measured by the condensate fraction
$n_0$, which is estimated through the asymptotic behavior of the one-body
density matrix $\rho (r)/\rho$, namely $n_{0} = \lim_{r \to \infty} \rho (r)/\rho$. 
The one-body matrix is an operator which is non-diagonal in coordinate 
space and does not commute with the Hamiltonian $H$ ($[H,\hat{\rho}]\neq 0$) 
so that DMC output for $n_{0}$ is a mixed estimator, though bias stemming from the trial wave function
can be reduced significantly by means of extrapolated estimator techniques.~\cite{cazorla08,cazorla08b}  

Vitali \emph{et al.}~\cite{shadow2} have recently
adapted the PIGS formalism to the symmetrized SWF to investigate 
strictly two-dimensional solid $^{4}$He with it.
Interestingly, the authors of this work conclude with the 
non-existence of ODLRO in perfect 2D solid $^{4}$He. We will comment on this
finding in the next paragraph in the light of our $\rho_{s} / \rho$ results.

\begin{figure}
\centerline{
        \includegraphics[width=0.8\linewidth,angle=0]{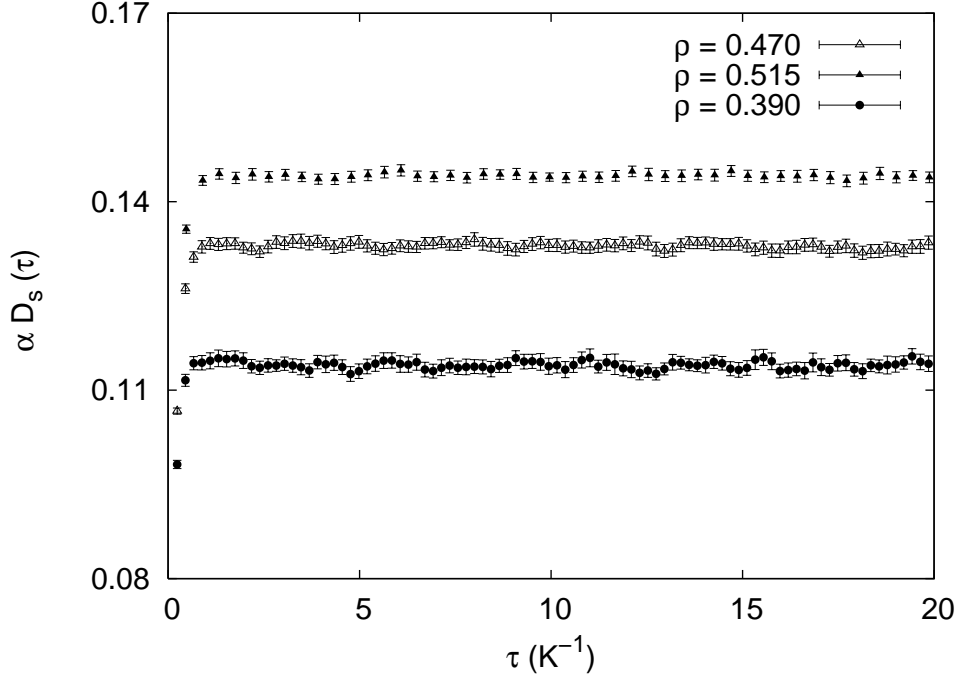}}%
        \caption{Diffusion of the center of mass of two-dimensional 
	   solid $^{4}$He in imaginary time calculated at a series of densities
	   near, below and above the melting point. A small upwards shift has 
	   been applied to the curves calculated at higher densities
	   in order to appreciate details of their slope.}
\label{fig:superfluid} 
\end{figure}

Differently to the estimation of $n_{0}$,  
the superfluid density of a bosonic system can be calculated exactly 
using DMC (whereas this has not been possible yet within the PIGS method) 
by extending the winding-number technique,
originally developed for PIMC calculations, to zero temperature.~\cite{zhang} 
Specifically, the expression of the superfluid fraction reads
\begin{equation}
\frac{\rho_{s}}{\rho} = \lim_{\tau \to \infty} \alpha \left(\frac{D_{s} (\tau)}{\tau} \right)~,
\label{eq:rhos}
\end{equation}
where $\alpha= N/ 4 D_0$ (two-dimensional case) with $D_0= \hbar^2/2m$,  
$D_{s} (\tau)=\langle({\bf R}_{\rm CM}(\tau) - {\bf R}_{\rm CM}(0))^2 \rangle$ and 
$\bf{R}_{\rm CM}$ is the center of mass of the particles in the plane. 
In Figure \ref{fig:superfluid}, we plot the function $D_{s} (\tau)$ calculated in the solid 
film at three different densities located near, above and below the corresponding freezing density. 
According to Eq.~(\ref{eq:rhos}), the superfluid fraction 
$\rho_{s} / \rho$ can be estimated directly from the slope of $D_{s} (\tau)$
at large imaginary time. 
In all the studied cases we find that the superfluid fraction of perfect 
two-dimensional solid $^{4}$He is vanishingly small, or to be more exact, it lies below our 
numerical threshold of $\sim 10^{-5}$. 
We found analogous $\rho_{s} / \rho$ results in the perfect three-dimensional case.~\cite{cazorla08}
As noted before, Vitali \emph{et al.}~\cite{shadow2} have recently studied perfect 2D solid $^{4}$He 
and concluded with the non-existence of ODLRO on this system. 

Very recently, we have studied strictly 2D solid
H$_{2}$ at zero-temperature with analogous approaches to the one used here.~\cite{cazorla08b} 
In two-dimensional molecular hydrogen, we found that in the regime of very low densities 
(negative pressure regime, $\rho_{\rm H_{2}} <  0.390~\sigma^{-2}$) 
a finite superfluid fraction appears in the crystal. 
Motivated by this result, and to understand the relation between normal 
and superfluid densities, we have carried out analogous calculations in 
2D solid helium at stable and metastable conditions (that is, at densities
above and below $\rho_{l} = 0.492~\sigma^{-2}$, respectively). It is worth noticing that
the DMC method has already been used to study ground-state properties of
metastable liquid $^{4}$He (overpressurized).~\cite{vranjes05}  
In the present case, we find a null value of $\rho_{s} / \rho$ for any density 
down to $\rho = 0.390~\sigma^{-2}$ (see Figure~\ref{fig:superfluid}). 
This result seems to be at odds with our previous findings in H$_{2}$ since solid helium possesses 
larger degree of quantumness compared to solid hydrogen. 
A possible explanation for this is that the density $\rho = 0.390~\sigma^{-2}$
is still far from the spinodal point of two-dimensional solid helium.
In fact, $^{4}$He is more compressible than H$_{2}$ (that is, 
$\mid \partial P/\partial V \mid _{\rm {He}} < \mid \partial P / \partial V \mid _{\rm {H_{2}}}$)     
so it is likely that the critical density at which mechanical instabilities appear in the first
system is below that of the second.
Also it must be noted that very dilute solid helium films are far from being realizable since 
the liquid phase is always energetically more favourable at densities below $\rho = 0.480~\sigma^{-2}$
(contrarily to what occurs in two-dimensional H$_{2}$).
In order to complete our study of low-dimensional solid helium
we have explored a system which can be considered as
somewhat more realistic, namely a quasi two-dimensional film.

\subsection{Quasi 2D solid $^{4}$He}

\begin{figure}
\centerline{
        \includegraphics[width=0.80\linewidth,angle=0]{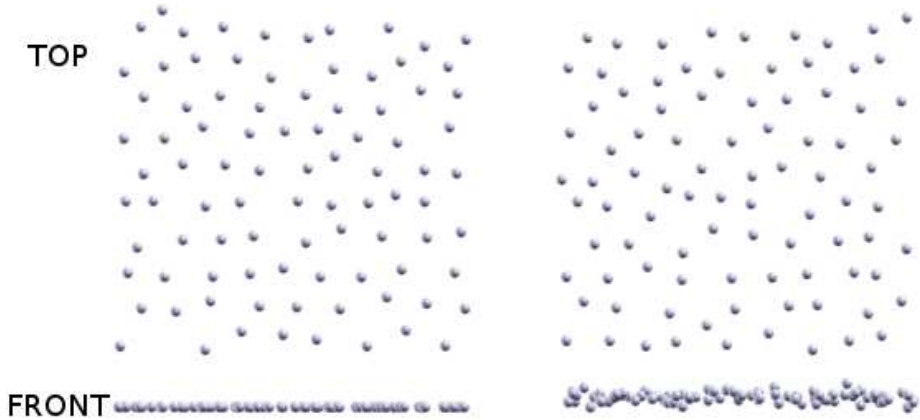}}%
        \caption{Top and front views of strictly two-dimensional (left) and quasi two-dimensional 
                 (right) solid $^{4}$He at density $\rho = 0.515~\sigma^{-2}$ and zero-temperature.}
\label{fig:2d-quasi2d}
\end{figure}

Very recent torsional oscillator-like experiments performed in the second 
layer of solid $^{4}$He adsorbed on graphite seem to point towards the 
possible existence of a new kind of supersolid phase.~\cite{saunders09}
Physical quantities such as the density of particles, temperature and 
degree of corrugation with the substrate, appear to have an important
effect on the value of this supersolid-like signal.
Aimed at investigating on the origins of this manifested low-dimensional supersolidity,
which is totally absent in strictly two-dimensional solid $^{4}$He, 
we have studied an interesting kind of simple model~: a quasi 2D film.
In this work, a quasi two-dimensional solid refers to a system of interacting
$^{4}$He particles with atomic displacements mostly confined to a plane 
but which can spread over the $z$-axis due to zero-point oscillations
(see Figure~\ref{fig:2d-quasi2d}).   
This model grasps some essential features of helium layers adsorbed on carbon-based surfaces 
like graphene and graphite and could be relevant to describe surface effects in bulk crystals. 
There are several advantages in exploring this model system instead of performing 
more realistic simulations of helium films adsorbed on carbon-based surfaces.~\cite{gordillo09,pierce99,greywall91,corboz08} 
First, is the reduction of computational cost which derives from ignoring
explicit interactions with the substrate and that allows us to explore 
the superfluid properties of the model upon a wide range of conditions. 
Second, confinement in the $z$-direction can be tuned at wish in order  
to explore the relation between the superfluid fraction $\rho_{s} / \rho$ and
magnitude of the spatial out-plane fluctuations 
$\langle \Delta z^{2} \rangle = \langle \left( z - \langle z \rangle \right)^{2} \rangle$,
which are related to the strength of the helium film interactions
with the substrate.
Certainly, realistic simulations of $^{4}$He films are necessary to 
fully understand the origins of supersolid manifestations, however, here
we assume simplified interatomic interactions and atomic structure (only the triangular lattice
is considered) in exchange for analyzing possible effects deriving from 
the density of particles and strength of the film-substrate interactions. 
The density range in which we concentrate corresponds to that of 
non-complete first layer of solid helium adsorbed on graphite 
(that is $0.416 < \rho < 0.745$~atom/$\sigma^{3}$)~\cite{corboz08}, so effects relating to promotion  
of atoms to second or higher layers are not considered. 

\begin{figure}
\centerline{
        \includegraphics[width=0.8\linewidth,angle=0]{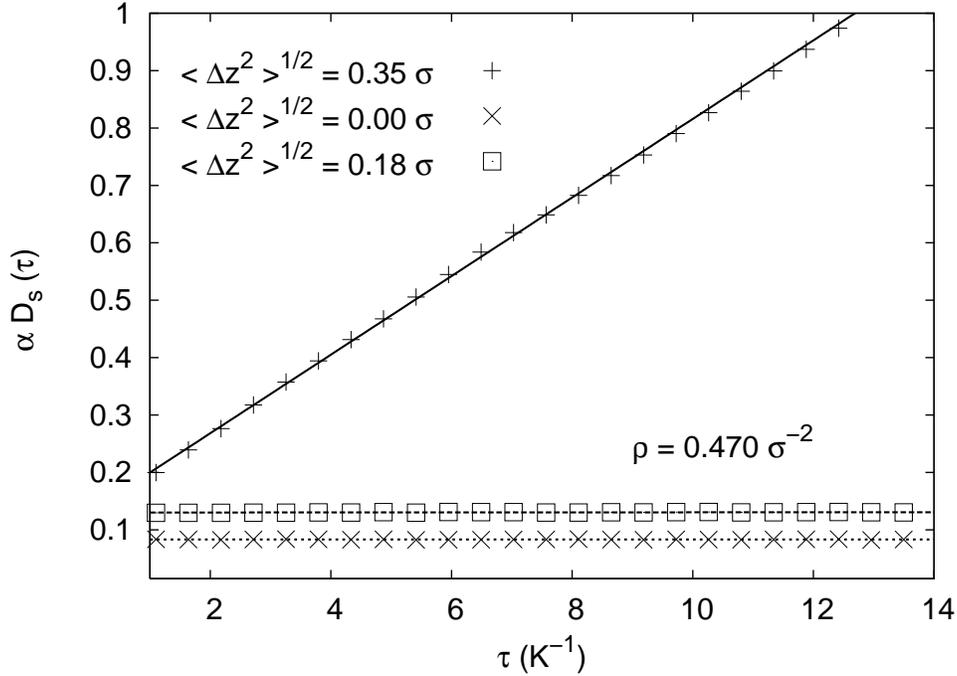}}%
        \caption{Diffusion of the center of mass of quasi two-dimensional 
		solid $^{4}$He calculated at the density $\rho = 0.470~\sigma^{-2}$ and  
		different values of the atomic $z$-spatial fluctuations $\langle \Delta z^{2} \rangle$.}
\label{fig:superquasi}
\end{figure} 

Local fluctuations of the atomic positions in the $z$-direction are achieved 
by imposing an external out-of-plane harmonic potential trap $V_{c} (z) \propto \left( z-z_{0}\right)^{2}$,
where $z_{0}$ is the equilibrium position of the film in that direction.
The Hamiltonian describing the quasi 2D system then is $H = T + V_{AzizII} + V_{c}$~.
The symmetrized trial wave function that we use to describe the
quasi two-dimensional system is
\begin{equation}
\Psi_{\rm SNJ}^{\rm{q-2D}}({\bf r}_1,\ldots,{\bf r}_N) = \prod_{i<j}^{N} f(r_{ij})~ 
\prod_{J=1}^{N} \left( \sum_{i=1}^{N} g(r_{iJ}^{xy}) \right)~ 
\prod_{i=1}^{N}\exp{\left( -\frac{1}{2} \chi z_{i}^{2} \right)}
\label{eq:3Dtwf}
\end{equation}
where $r_{iJ}^{xy}$ is the projection of the vector
${\bf r_{i}} - {\bf R_{J}}$ over the $xy$-plane, lattice vectors being
$\lbrace {\bf R_{J}} = (a, b, 0)\rbrace$, and the value of parameter
$\chi$ is explicitly related to the value of the
atomic spatial $z$-fluctuation by $\langle \Delta z^{2} \rangle = 1 / 2\chi$~.
Trial wave function $\Psi_{\rm SNJ}^{\rm{q-2D}}$ is equivalent to  
$\Psi_{\rm SNJ}$ in Eq.~(\ref{snjtrial}) but with additional  
Gaussian localizing factors on the $z$-direction; these localizing functions correspond to the 
exact Schr\"odinger equation solution of a particle moving under the action of the harmonic 
potential field $V_{c}$. Since the computational technique used in this section is the DMC 
method and the magnitude of the atomic $z$-spatial fluctuations analyzed is fairly small, 
we have not attempted to construct explicit $z$-pairwise correlations on the trial 
wave function. It must be noted that these correlations are implicitly taken into account by 
the Jastrow factor contained in $\Psi_{\rm SNJ}^{\rm{q-2D}}$.    
Variational parameters contained in expression (\ref{eq:3Dtwf}) are set to the
same value than used in the study of strictly two-dimensional solid
$^{4}$He, made the exception of $\chi$ which is varied according to the value
of the spring constant corresponding to the harmonic trap.

\begin{figure}[t]
\centerline{
        \includegraphics[width=0.8\linewidth,angle=0]{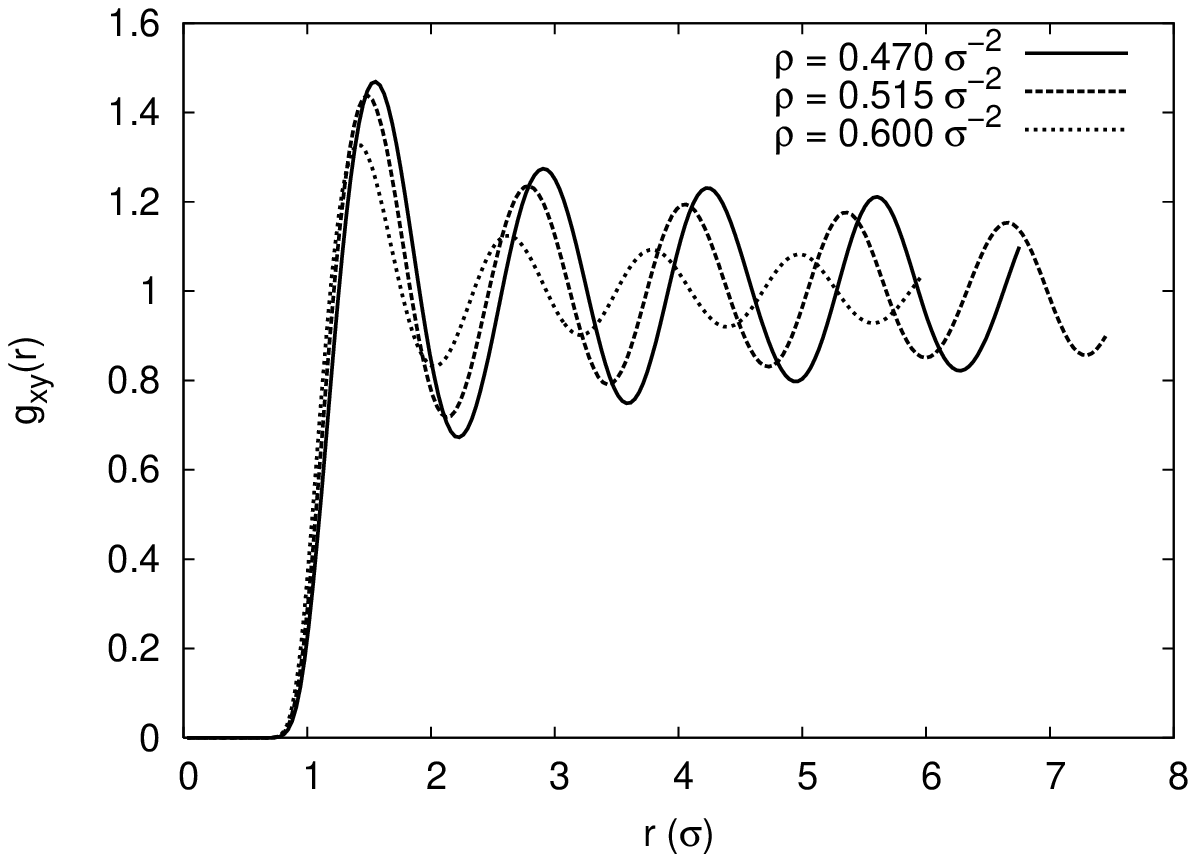}}%
        \caption{Radial pair-distribution function calculated on the $xy$-plane of the
		 quasi 2D film at different densities and fixed amplitude of $z$-fluctuations 
		 $\langle \Delta z^{2} \rangle^{1/2} = 0.35~\sigma$~. At density
		 $\rho = 0.600~\sigma^{-2}$, the peaks and valleys of the $g_{xy}(r)$ function
		 turn out to be appreciably less pronounced than in the other cases considered.}
\label{fig:grquasi3D}
\end{figure}

\begin{table}
\begin{center}
\begin{tabular}{c||c|c|c}
$  $  & \multicolumn{3}{c}{$\langle \Delta z^{2} \rangle^{\frac{1}{2}}~(\sigma)$}\\
\hline
$ \rho~(\sigma^{-2}) $ &  $\quad  0.18 \quad$  & $\quad 0.22 \quad$  &  $\quad 0.35  \quad$  \\
\hline
$ 0.470 $   &  $ 0.003(1)  $  & $ 0.16(1) $  &  $ 6.85(1)  $  \\
$ 0.515 $   &  $ 0.015(1)  $  & $ 0.010(1)$  &  $ 17.27(1) $  \\
$ 0.600 $   &  $ 0.0       $  & $ 0.009(1)$  &  $ 54.21(1) $  \\
\hline
\hline
\end{tabular} 
\end{center}
\caption{Calculated superfluid fraction (expressed in $\%$) of quasi-2D solid $^{4}$He
as a function of the density of particles and atomic spatial $z$-fluctuation $\langle \Delta z^{2} \rangle$~.  
 }
\label{tab:super3D}
\end{table}

\begin{figure}[t]
\centerline{\includegraphics[width=0.7\linewidth,angle=0]{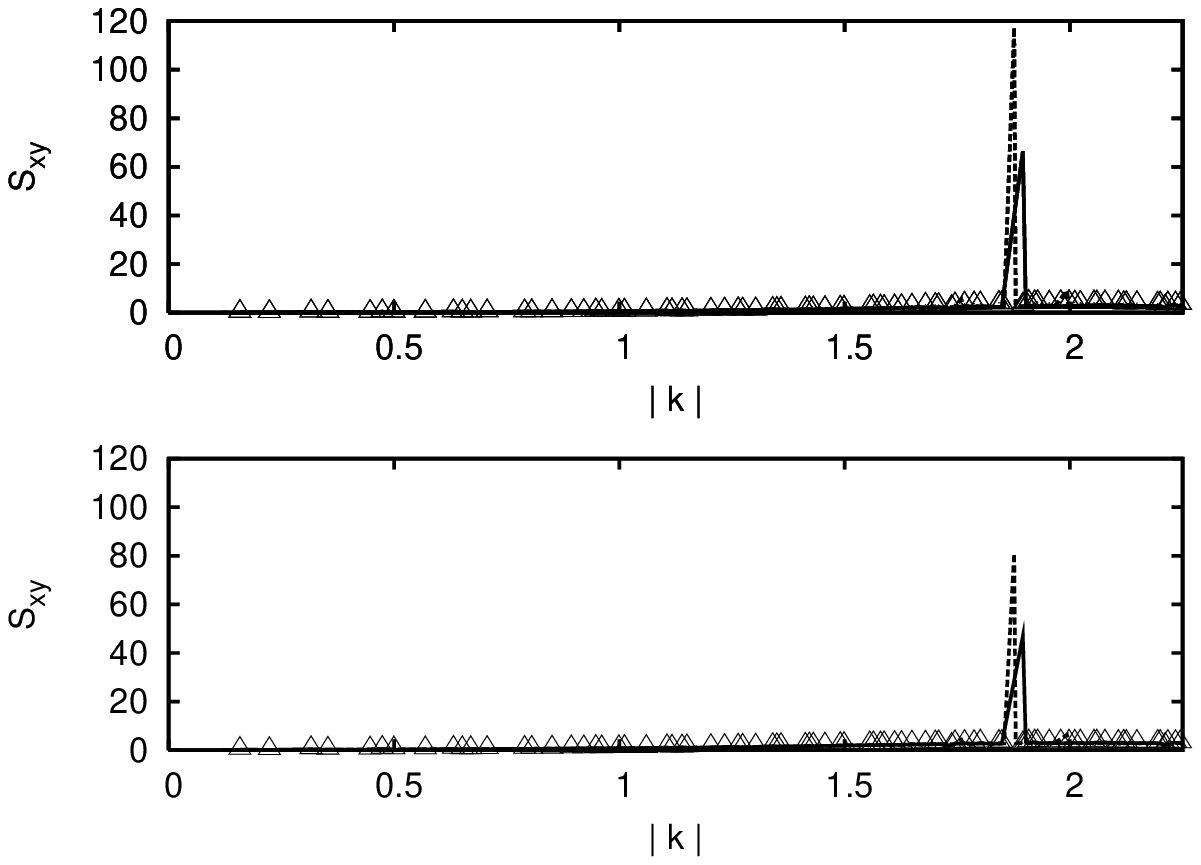}}%
        \caption{Calculated radial averaged structure factor $S_{xy} (k)$ 
                 of the quasi 2D solid film at $\rho = 0.515~\sigma^{-2}$
                 ($k$ is in units of $\sigma^{-1}$).
                 Solid lines represent calculations performed with $N = 120$ atoms,  
                 dashed lines calculations performed with $N = 224$ atoms and triangles
                 correspond to results obtained for a quasi 2D liquid system ($N = 120$ atoms).  
                 In the lower and upper panels, we show results obtained for $z$-confinement    
                 $\langle \Delta z^{2} \rangle^{1/2} = 0.35~\sigma$ and  
                 $\langle \Delta z^{2} \rangle^{1/2} = 0.22~\sigma$, respectively.}
\label{fig:skquasi3D}
\end{figure}

Next, we comment on the superfluid fraction results obtained at several densities and
$z$-confinement conditions. 
In Table~I, we report the dependence of the estimated superfluid density fraction on the 
density of particles and the $z$-coordinate fluctuation $\langle \Delta z^{2} \rangle$;
in Figure~\ref{fig:superquasi} we also show the evolution of function $D_{s} (\tau)$ on 
imaginary time at density $\rho = 0.470~\sigma^{-2}$ and several atomic $z$-confinements.
It is found that in the $\langle \Delta z^{2} \rangle^{1/2} \le 0.08~\sigma$ cases (not shown) 
the superfluid density is always vanishing, which turns 
out to be consistent to what it is found in the strictly two-dimensional case.  
On the contrary, when confinement on the $z$-direction is shallow, 
that is case $\langle \Delta z^{2} \rangle^{1/2} = 0.35~\sigma$, the value of $\rho_{s} / \rho$
is always large and increases as the density of particles is raised (see Table~I). 
Results in the last column of Table~I appear to show a competition 
between in-plane  and out-plane interactions;
as the system is compressed, in-plane repulsive interactions become progressively more important 
so that atoms prefer to spread over the $z$-direction wherein potential confinement is mild and they 
can move more freely. This has the overall effect of enhancing the superfluid response of the 
system. In the $\langle \Delta z^{2} \rangle^{1/2} = 0.18$ and $0.22~\sigma$ cases,
estimated $\rho_{s} / \rho$ trends on density are not that monotonic. 
At $\langle \Delta z^{2} \rangle^{1/2} = 0.18~\sigma$, we see 
that $\rho_{s} / \rho$ first increases when the density of particles is 
raised whereas next it diminishes down to zero-value under further compression 
of the system. This behavior can be understood in terms of the imposed $z$-axis confinement 
and atomic in-plane interactions as well.
When the density of particles is first increased, atoms minimize their potential energy 
by spanning over the $z$-axis in order to keep a distance from their neighbors and
move as freely as possible. The total space available for the atoms then becomes larger 
so the effective density of the system becomes smaller. 
The value of $\rho_{s} / \rho$ consequently increases.
However, when density is further raised, out-plane atomic excursions are not  
favourable any more because large displacements along the $z$-direction require too much 
energy. The value of $\rho_{s} / \rho$ then decreases because atomic motion is tightly 
bound to the $xy$-plane and the effective density of the system becomes large. 
The competition between in- and out-plane interactions as modulated by 
the density of particles is responsible for the enhancement/depletion of the
observed superfluid response of the system.    
At $z$-axis confinement $\langle \Delta z^{2} \rangle^{1/2} = 0.22~\sigma$, the trend of $\rho_{s} / \rho$ 
on density exhibits an intermediate behavior between that found in the $0.18$ and $0.35~\sigma$
cases. 

Regarding the stability of the quasi 2D solid film, 
we note that in all the studied cases crystal-like order has
been found as witnessed by (i)~marked oscillating shape of the 
radial pair-distribution functions $g_{xy}(r)$ obtained considering 
the projection of the atomic positions on the $xy$-plane (see Figure~\ref{fig:grquasi3D}), 
and (ii)~peaked pattern of the corresponding radial averaged 
structure factors $S_{xy}$ that scale with the number of atoms 
(see Figure~\ref{fig:skquasi3D}).
Nevertheless, in the $\langle \Delta z^{2} \rangle^{1/2} = 0.35~\sigma$ cases the film is 
likely to be a kind of glass system since the value of the corresponding maximum 
$S_{xy} (k)$ peaks are small 
(see Figure~\ref{fig:skquasi3D}, lower panel) 
and the $\rho_{s} / \rho$ values obtained are quite large (last column in Table~I)
in comparison to the results obtained upon tighter $z$-confinement.  
In fact, large superfluid fractions ($\sim 10-60~\%$) have been estimated
in metastable $^{4}$He glass systems using the PIMC method.~\cite{glass} 
Moreover, we have calculated the total energy of a quasi 2D liquid system 
at same density and $\langle \Delta z^{2} \rangle^{1/2}$ conditions than in the 
quasi 2D solid system, using a Jastrow factor and Gaussian $z$-localizing factors 
as importance sampling. It is found that in all the
$\langle \Delta z^{2} \rangle^{1/2} = 0.35~\sigma$ cases the quasi-2D 
solid film is metastable (that is, $E^{\rm q-2D}_{\rm J} < E^{\rm q-2D}_{\rm SNJ}$)
whereas is stable in the rest of $\langle \Delta z^{2} \rangle^{1/2}$ 
and $\rho$ cases (that is, $E^{\rm q-2D}_{\rm SNJ} < E^{\rm q-2D}_{\rm J}$).
For instance, at $\langle \Delta z^{2} \rangle^{1/2} = 0.35~\sigma$
and $\rho = 0.515~\sigma^{-2}$ we obtain $E^{\rm q-2D}_{\rm J} = 2.71(2)$~K/atom 
and $ E^{\rm q-2D}_{\rm SNJ} = 3.66(2)$~K/atom, while at 
$\langle \Delta z^{2} \rangle^{1/2} = 0.18~\sigma$
and same density we obtain $E^{\rm q-2D}_{\rm J} = 16.29(3)$~K/atom
and $ E^{\rm q-2D}_{\rm SNJ} = 15.81(2)$~K/atom. 
This last outcome, in analogy with the 3D case, appears to corroborate 
the hypothesis that the results obtained at $z$-confinement 
$\langle \Delta z^{2} \rangle^{1/2} = 0.35~\sigma$  
correspond to quasi 2D glass systems. 

\begin{figure}[t]
\centerline{
        \includegraphics[width=0.8\linewidth,angle=0]{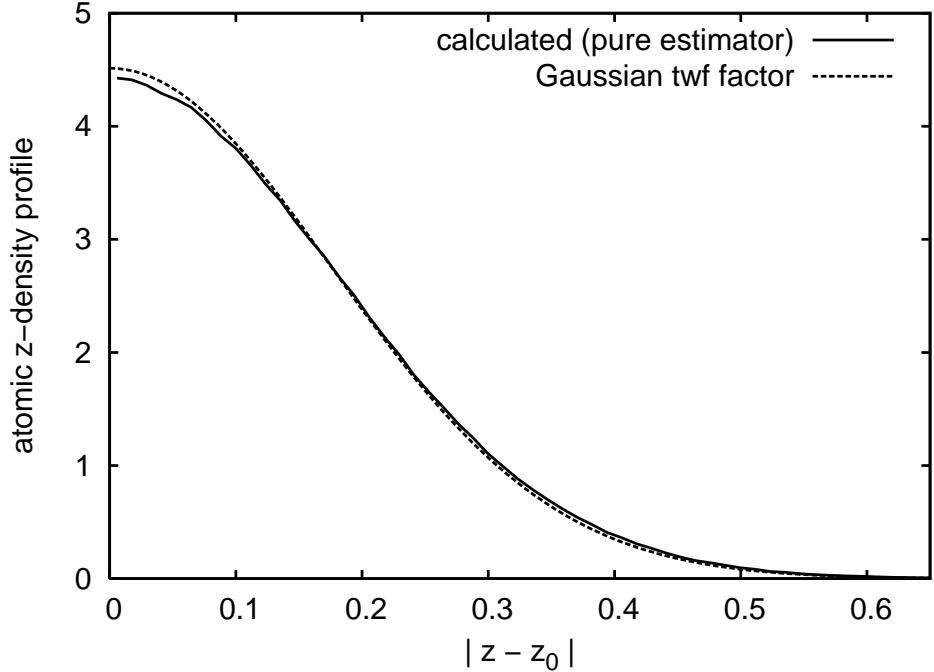}}%
        \caption{Calculated atomic $z$-density profile (solid line)    
                 in the quasi 2D solid system at $\rho = 0.470~\sigma^{-2}$ and
                 $\langle \Delta z^{2} \rangle^{1/2} = 0.18~\sigma$. For comparison,
                 we plot the corresponding normalized Gaussian $z$-localizing factor
                 ($\chi = 16~\sigma^{-2}$) entering trial wave function $\Psi_{\rm SNJ}^{\rm{q-2D}}$.
                 Distance is in units of $\sigma$.}
\label{fig:zprofile}
\end{figure}

It is worth noticing that although particles in the quasi 2D film are allowed to move in the $z$-direction,
there is not enough energy to excite the levels of the transverse confinement
and the system is kinematically two-dimensional. The radial motion is frozen 
to zero-point oscillations and the magnitude of $z$-fluctuations therefore is related to 
the length $\langle \Delta z^{2} \rangle^{1/2}$.
We illustrate this in Figure~\ref{fig:zprofile} where we plot the atomic 
$z$-density profile calculated in the quasi 2D solid film at $\rho = 0.470~\sigma^{-2}$ and
$\langle \Delta z^{2} \rangle^{1/2} = 0.18~\sigma$ using the pure estimators
technique,~\cite{casulleras} and compare it with the corresponding normalized Gaussian 
$z$-localizing factor entering $\Psi^{\rm q-2D}_{\rm SNJ}$ (case $\chi = 16~\sigma^{-2}$).

\section{Conclusions}
\label{sec:discussion}

We have studied two- and quasi two-dimensional solid $^{4}$He at zero temperature 
by means of the diffusion Monte Carlo method and using the recently proposed symmetrized 
trial wave function $\Psi_{\rm SNJ}$ as importance sampling. 
We have estimated the superfluid density fraction of two-dimensional solid $^{4}$He
at $T = 0$ and found that is negligible down to a density of $\rho = 0.390~\sigma^{-2}$.
Importantly, by allowing the atoms to move along the $z$-axis we observe 
the appearance of a superfluid response that coexists with the crystalline order of the
system. The magnitude of this response is shown to depend on the degree of $z$-axis 
confinement and the density of particles.  
This finding is valuable for the realization and interpretation of more realistic
simulations of helium layers adsorbed on carbon-based surfaces, where the interactions 
with the substrate must be taken into account accurately in order  
to make rigorous judgements about the existence of superfluidity and/or 
ODLRO. In view of the present results for the quasi two-dimensional solid,  
it can be suggested that a well-suited system where to observe a finite superfluid
signal is the first layer of $^{4}$He on top of graphene or graphite~\cite{gordillo09}
which stabilizes in a triangular lattice and possesses relatively small density. 
Work to verify this hypothesis is in progress.

\section*{Acknowledgements}
We acknowledge partial financial support from DGI (Spain) Grant No.
FIS2008-04403 and Generalitat de Catalunya Grant No. 2009SGR-1003.

\end{document}